\documentclass[prl,showpacs,aps,epsfigure,twocolumn]{revtex4-1}

\usepackage{array}
\usepackage{amsmath}
\usepackage{graphicx}
\usepackage{dcolumn}
\usepackage{bm}
\usepackage{graphicx}
\usepackage{amsmath}
\usepackage{latexsym}
\usepackage{amsfonts}
\usepackage{amssymb}
\usepackage{array}
\usepackage{epsfig}
\usepackage{epstopdf}
\usepackage{times}

\begin{document}

\title{Photon production from the vacuum close to the super-radiant transition:\\When Casimir meets Kibble-Zurek}
\author{G. Vacanti$^1$, S. Pugnetti$^2$, N. Didier$^2$, M. Paternostro$^3$, G. M. Palma$^4$, R. Fazio$^2$$^,$$^1$, V. Vedral$^1$$^,$$^5$$^,$$^6$ }
\affiliation{$^1$Center for Quantum Technologies, National University of Singapore, 1 Science Drive 2, Singapore\\
$^2$NEST, Scuola Normale Superiore and Istituto Nanoscienze-CNR, I-56126 Pisa, Italy.\\
$^3$School of Mathematics and Physics, Queen's University, Belfast BT7 1NN, UK\\
$^4$NEST-CNR (INFM) and Dipartimento di Fisica, Univerisita' degli Studi di Palermo, Via Archirafi 36, Palermo, Italy\\
$^5$Clarendon Laboratory, University of Oxford, Parks Road,  Oxford 0X1 3PU, UK \\
$^6$Department of Physics, National University of Singapore, 3 Science Drive 4, Singapore}

\date{\today}

\begin{abstract}
The dynamical Casimir effect (DCE) predicts the generation of photons from the vacuum due to the parametric amplification of the quantum fluctuations of an electromagnetic field. The verification of such effect is still elusive in optical systems due to the very demanding requirements of its experimental implementation. This typically requires very fast changes of the boundary conditions of the problem. We show that an ensemble of two-level atoms collectively coupled to the electromagnetic field of a cavity (thus embodying the quantum Dicke model), driven at low frequencies and close to a quantum phase transition, stimulates the production of photons from the vacuum. This paves the way to an effective simulation of the DCE through a mechanism that has recently found experimental demonstration. The spectral properties of the emitted radiation reflect the critical nature of the system and allow us to link the detection of DCE to the Kibble-Zurek mechanism for the production of defects when crossing a continuous phase transition. 
\end{abstract}


\maketitle

When $N$ two-level atoms interact collectively with a single mode of the electromagnetic field inside a cavity, thus realizing the so-called Dicke model~\cite{dicke}, a critical value of the atom-photon coupling $g_c$ exists at which the system undergoes a quantum phase transition, generally referred to as the {\em super-radiant transition}. Below such coupling, the atoms are in their ground state and the cavity field is unpopulated.  Conversely, above $g_c$ there is a spontaneous symmetry breaking and the photon field gets populated through a mechanism producing a displaced coherent state~\cite{barnett}. The experimental demonstration of the super-radiant transition in the Dicke model has remained outstanding until recently, when a key result has been achieved in a set-up involving intra-cavity Bose-Einstein condensates~\cite{esslinger}. A super-radiant transition has been enforced by exploiting the spatial self-organization of the atoms in an intra-cavity condensate coupled to the cavity field and subjected to an optical-lattice potential. 

Here we investigate the relation between equilibrium and dynamical properties of a Dicke system brought close to a quantum phase transition. We prove that, at the super-radiant transition, a DCE-like mechanism~\cite{casimir1,casimir2} arises from the use of a time-dependent driving and results in a flux of photons generated from the vacuum fluctuations. DCE has been predicted to occur in QED settings involving a cavity with oscillating end mirror~\cite{dodonovreview}. This scheme, however, appears to be technologically demanding given the prohibitively large frequency at which the mirrors should vibrate to produce a measurable flux of  photons. Notwithstanding some interesting proposals~\cite{nori,dodonov,deliberato} having the potential to ease the requirements for its observability, an experimental demonstration of DCE is still elusive in the optical domain. Recently, a DCE-like mechanism has been observed in an experiment performed using microwaves~\cite{noriDCE}.

Our proposal pursues a different direction: we observe that, on approaching the Dicke super-radiant phase transition, the frequencies at which the DCE-like effect becomes observable are lowered, thus narrowing the gap separating the experimental state-of-the-art from the observation of the effect. Moreover, we unveil an intriguing connection between the occurrence of DCE through the mechanism we propose and the Kibble-Zurek mechanism (KZM)~\cite{KZ1,KZ2}. The latter predicts the formation of defects in a quantum many-body system dragged through a critical point~\cite{dziarmaga,zurek05,polkovnikov05} and is due to the inability of the system to remain in its ground state. The production of defects occurs regardless of how slowly the dragging is performed and the mechanism has been shown to be related to adiabatic quantum computation~\cite{farhi01} and quantum annealing~\cite{santoro02}. We are thus able to bridge two fundamental phenomena in out-of-equilibrium quantum systems with the goal of simplifying their observation. The recent demonstration of the Dicke super-radiant transition~\cite{esslinger}, which is the building block of our proposal, marks a promising starting point towards an experimental investigation along the lines of our work. 

\begin{figure}[t]
\begin{tabular}{cc}
\includegraphics[width=.49\columnwidth]{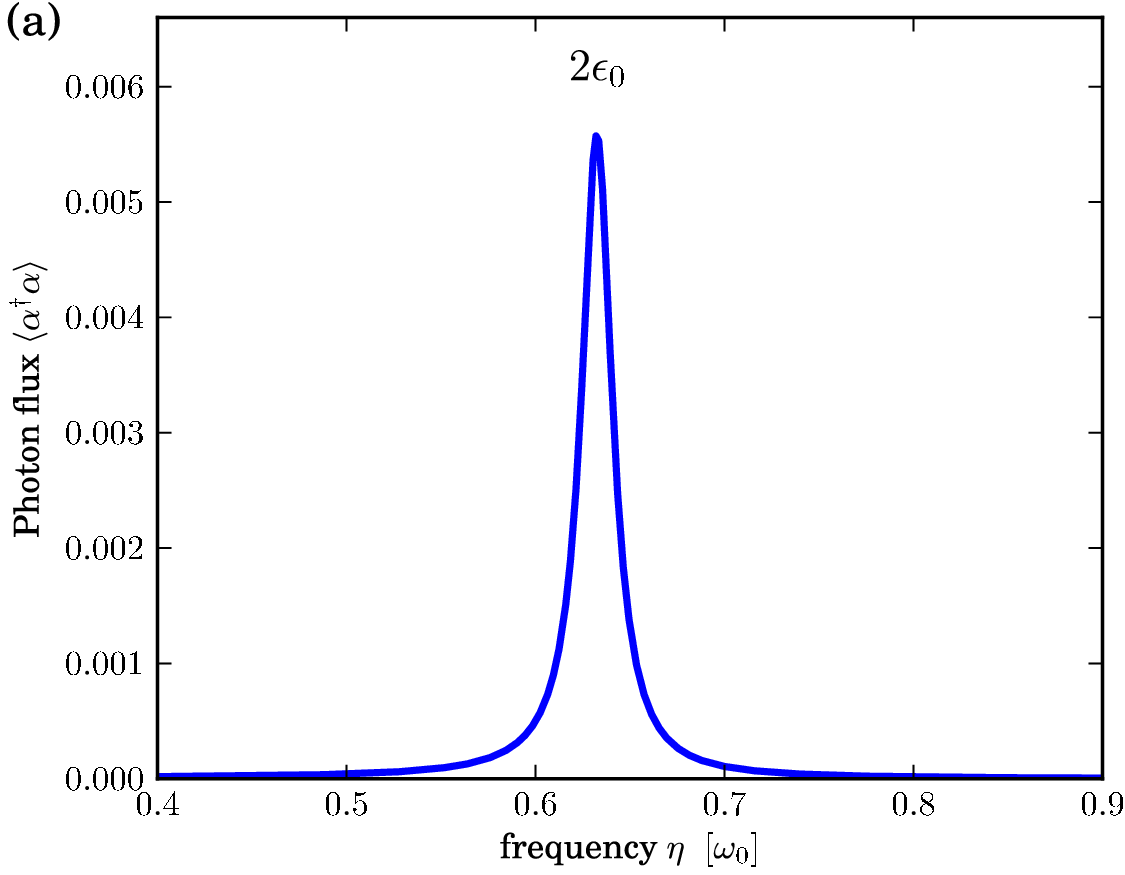}&
\includegraphics[width=.49\columnwidth]{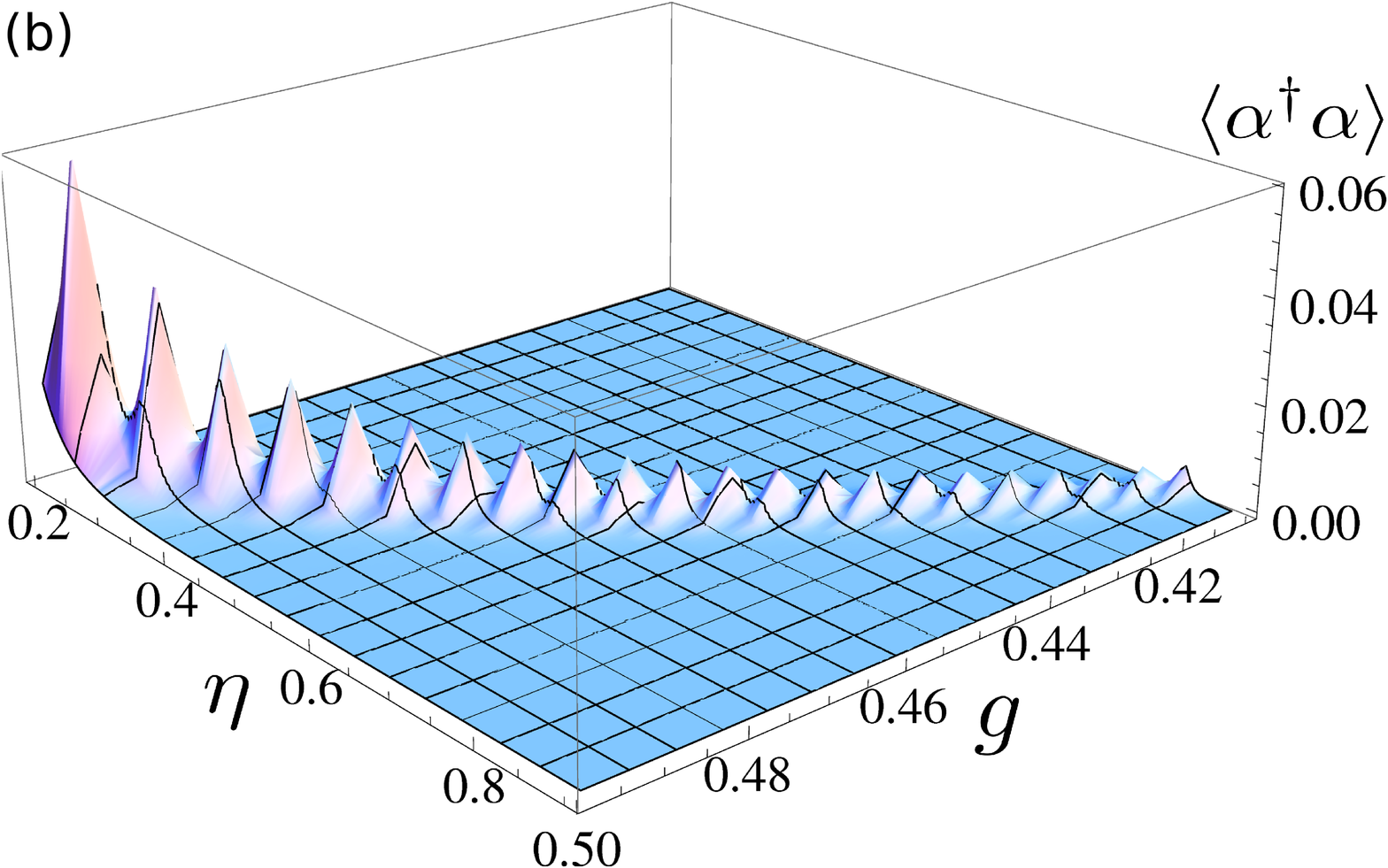}
\end{tabular}
\caption{{Radiation flux outside the cavity.} (a) Flux of photons outside the cavity against $\eta$ for $g /\omega_a= 0.9,$ $g_c /\omega_a=0.45$, $\gamma/\omega_a = 0.005$, and $\lambda/\omega_a = 0.005$. For these parameters, $\epsilon_0 /\omega_a\approx 0.315$. (b) Flux of photons outside the cavity against $\eta$ and $g$ for $\omega_b/\omega_a=1$, $\gamma/\omega_a=0.005$ and $\lambda/\omega_a=0.005.$ }
\label{phout}
\end{figure}
We consider a cold system of $N$ two-level atoms, collectively interacting with the field of a single-mode cavity whose annihilation (creation) operator is $\hat{a}$ ($\hat{a}^\dag$). Each two-level atom is modelled as a pseudo-spin whose Pauli spin matrices are $\{\hat{\sigma}_\pm^i, \hat{\sigma}_z^i\}$ ($i{=}1,..,N$). We realistically assume a small-sized atomic sample, neglect the variations of the cavity field at its location and take the coupling strength as uniform. Then, we introduce the total angular momentum $\hat{J}$ of the atomic sample with components $\hat{J}_\pm=\sum_i{\hat{\sigma}_\pm^i}$ and $\hat{J}_z=\sum_i{\hat{\sigma}_z^i}$ and consider time-dependent single-spin energy splittings $\omega_b(t)$. The Hamiltonian of the system in the dipole approximation thus reads (we set $\hbar=1$ throughout the paper)
\begin{equation}
\hat  H_0= \omega_a \hat{a}^\dag \hat{a} + \omega_b(t) \hat{J}_z + \frac{g}{\sqrt{2j}} (\hat{a}^\dag + \hat{a}) (\hat{J}_+ + \hat{J}_-),
\label{H0j}
\end{equation}
where $\omega_a$ is the frequency of the cavity and $g$ is the atom-field coupling constant. 
Analogously to Refs.~\cite{dodonov,deliberato}, we take $\omega_b(t) = \omega_0 + \lambda \sin(\eta t)$. The parameter $j$ is the {\it cooperation number} in the Dicke theory, that is an eigenvalue of $\hat{J}^2$ which, together with the eigenstates of $\hat J_z$, is used to build the Dicke states. The ensemble of $N$ two-level atoms is then described as a pseudo-spin of size $j=N/2$. In this case, photons are generated at $\eta_\text{res} = 2 \omega_a$. In Ref.~\cite{Kirilyuk} a model similar to ours but based on a semi-classical approach has been addressed to relate DCE-like effects to Dicke super-radiance. Here we perform a full quantum treatment of both the atom-light interaction and the effects on the photon statistics induced by the driving of the atomic subsystem. Moreover, as discussed in the second part of the paper, we will unveil the connection between the DCE-like effects and the KZ mechanism.

{Eq.~(\ref{H0j}) strongly resembles the many-body Landau-Zener problem studied in Refs.~\cite{manybodyLZ1,manybodyLZ2}. A crucial difference between the two cases is the presence of the counter-rotating terms in Eq.~(\ref{H0j}). These lead {\em both} to the super-radiant transition and the production of photons}. {For $N\gg{1}$, the Holstein-Primakoff representation} of the angular momentum~\cite{HP} can be used to approximate the atomic cloud to a non-linear harmonic oscillator. However, for a large atomic sample (as in the case here), we can take $\hat{J}_+\approx \sqrt{2 j} \hat{b}^\dag$ with $[\hat b,\hat b^\dag]=1$ and retain the harmonic approximation. Eq.~(\ref{H0j}) thus reduces to 
\begin{equation}
\label{Htab}
\hat   H= \omega_a \hat a^\dag \hat a + \omega_b(t) \hat b^\dag \hat b + g (\hat a^\dag + \hat a) (\hat b^\dag + \hat b),
\end{equation}
which  is easily diagonalised: the normal frequencies $\epsilon_{\pm}(t)$ and modes $q_\pm(t)$ are discussed in Appendix A (their form is not relevant for our purposes). The normal-mode description of Eq.~(\ref{Htab}) simplifies the analysis of the critical properties of the time-modulated Dicke model~\cite{emary}, which shows the existence of a critical value $g_c(t) = \sqrt{\omega_a\omega_b(t)}/2$ at which a phase transition occurs. In the phase corresponding to $g<g_c$, which we dub {\it normal phase}, the number of photons in the cavity mode is very small and $\langle \hat a \rangle=\langle \hat b \rangle=0$. Upon approaching $g_c(t)$, the number of photons increases and the system reaches the so-called {\it super-radiant phase} at $g{>}g_c(t)$, where the cavity mode is macroscopically populated even in the ground state of the system and $\langle \hat k \rangle\neq0~(k{=}a,b)$ due to spontaneous symmetry-breaking: while at small couplings the total number of excitations $\langle \hat a^\dag \hat a+\hat b^\dag \hat b\rangle$ is conserved, at the phase transition such a symmetry is broken, resulting in the population of the cavity field. Here, we will only refer to the normal phase.

Let us first address the case of a lossless evolution. For parameters of the system such that $\epsilon_+ (t) \gg\{\epsilon_-(t),\eta\}$, the non-critical mode $\hat{q}_+$ will not contribute to the photon production and we can consider only $\hat{q}_-$. To simplify the notation, we drop the index and set $\epsilon_-(t) \equiv \epsilon(t)$ from now on. Moreover, taking $\lambda$ small, the system can be treated as a harmonic oscillator with frequency $\epsilon_0$ perturbed by a weak driving at the modulation frequency $\eta$. In this framework, excitations can be created only at the resonance condition $\eta=k\epsilon_0$ ($k\in\mathbb{Z}$). If the system is initially in its ground state, which corresponds to the vacuum of the effective harmonic oscillator, such dynamics lead to the generation of photons from the vacuum inside the cavity, along the lines of DCE, already at $\eta = 2 \epsilon_0$. The number of photons generated increases as $g\rightarrow g_c$. This is very important: as $\epsilon_0\rightarrow0$ when the system approaches its critical point, the $\eta$ needed to observe the DCE-like effect is lowered, thus bringing its verification closer to experimental feasibility.

The assumption of unitarity is not realistic and we now include the leakage of photons from the cavity, a mechanism that is conveniently tackled by means of the input-output formalism for optical cavities~\cite{Walls} and by modelling the bath to which the cavity field is coupled as a distribution of harmonic oscillators with associated operators $(\hat \alpha_{\nu},\hat \alpha_{\nu}^\dag)$, having frequency $\nu$ and interacting with the cavity field with strength $k_\nu$ according to the Hamiltonian 
$V {=} i\int_{0}^{\infty} \mathrm k_{\nu} (\hat \alpha_{\nu} \hat a^\dag - \hat a \hat\alpha_{\nu}^\dag ){d} \nu$.
The method used to tackle such open dynamics is fully described in Appendix B. Here we only state that, by assuming a flat density of states in the bath $\rho(\nu)$ vanishing for $\nu{<}0$ and $k_\nu{=}k$ ($\forall\nu{>}0$), manageable expressions for the effective dissipation rates affecting the evolution of the system at hand are found. Moreover, we can determine the stationary mean number of photons inside the cavity $\langle a^\dag a \rangle= \lim_{t\rightarrow \infty} \langle a^\dag(t) a(t) \rangle$ and, from this, the out-coming photons flux $\langle\hat\alpha^\dag\hat\alpha\rangle= \int_{0}^{\infty} \mathrm{d}\omega \langle \alpha^{\mathrm{out}^\dag}_{\nu} \alpha^\mathrm{out}_{\nu} \rangle$. Here, $\hat\alpha^\text{out}_\nu$ ($\hat\alpha^\text{out\dag}_\nu$) is the field annihilation (creation) operator of the output mode at frequency $\nu$~\cite{CiutiCaru}. 

\begin{figure}[t]
\includegraphics[width=.8\columnwidth]{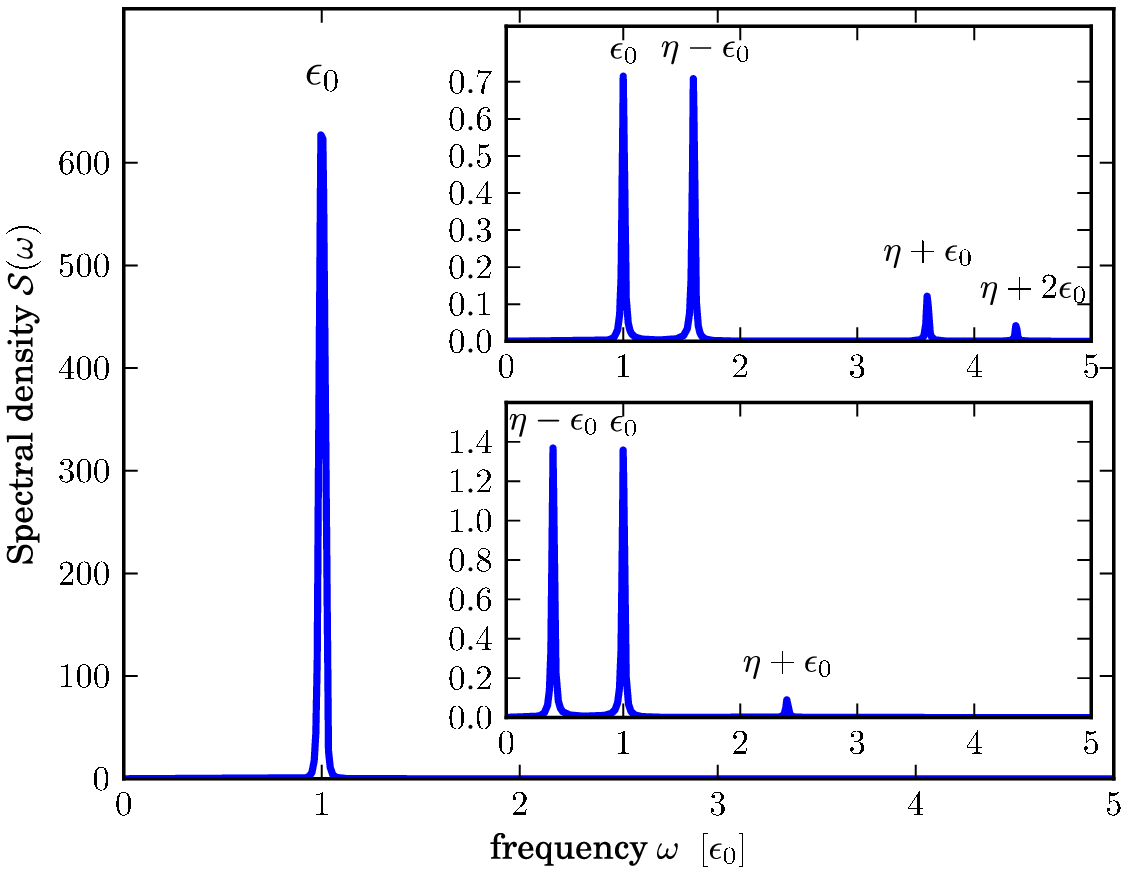}
\caption{{Spectral density of the output photons.} Taking  $\omega_a = \omega_b=1$, $\lambda = 0.005$, $\gamma=0.005$, $g=0.9$ and $g_c = 0.45,$ we find $ \epsilon_0 = 0.315$.  We have taken $\eta/2\epsilon_0 = 1$ (corresponding to resonance conditions,  main panel),  $\eta/2\epsilon_0 = 0.7$ (upper inset),  $\eta/2\epsilon_0 = 1.3$ (lower inset).}
\label{spectra}
\end{figure}
We can now discuss the qualitative features of the mechanism here achieved. First, we find no out-coming photon for an unmodulated driving field. On the contrary, any time-dependent modulation generates a constant flux of photons. This is clearly understood by taking the case of a small-amplitude modulation and stopping at the first order in $\lambda$~\cite{eisert}. In Fig.~\ref{phout}, we show the out-coming photon-flux for a damping rate $\gamma$ (positive and constant for $\omega{>}0$) against $\eta$ and $g$. In Fig.~\ref{phout}~(a) a resonance peak is clearly visible at $\eta \approx 0.63$ when $g=0.45$, corresponding to $g=0.9 g_c$ and $\epsilon_0\approx 0.315$, which confirms that a resonance peak is achieved at $\eta \approx 2 \epsilon_0$. Such a prediction is strengthened by Fig.~\ref{phout} (b), where the flux of out-coming photons is plotted against $g$ and $\eta$ and shows resonances at $\eta \approx 2 \epsilon_0$, regardless of the value of $\epsilon_0$. 

We complete our study by considering the spectral density $\mathcal{S}(\omega)=\langle\hat\alpha^{{\rm out}\dag}(\omega)\hat\alpha^{{\rm out}}(\omega)\rangle$ of the output field, which is  plotted in Fig.~\ref{spectra} for a weak modulation driving at variable $\eta$. When $\eta{=}\eta_{\rm res}{=}2 \epsilon_0$, the spectrum reveals a single sharp peak at $\omega \approx \epsilon_0$ [cf. Fig.~\ref{spectra}, main panel].  In the non resonant regime, the emission at $\omega \approx \epsilon_0$ is drastically reduced and sidebands of enhanced emission at $\omega \approx \eta\pm\epsilon_0$ and $\omega \approx \eta+2\epsilon_0$ appear (cf. insets of Fig.~\ref{spectra}). 

A series of remarks are due. First, we reassure on the validity of the Holstein-Primakoff approximation. In all our calculations of the out-coming photon-flux, the number of excitations in the atomic system is much smaller than the actual number of atoms in the sample, thus ensuring that our framework holds. Second, due to the small frequency of the photons generated when $g$ approaches the critical value (the emission frequency is $\omega \approx \epsilon_0$), thermal noise in the output signal, which has not been included in our study, may be significant. However, the generated photons can be detected also in the presence of strong background noise simply by using a cavity with semi-transparent mirrors. In this scheme, the cavity mode is coupled with two thermal baths and photons are allowed to enter/abandon the cavity from both sides. A the two baths are uncorrelated, the noise can be virtually eliminated by measuring the correlations between the output modes. 

\begin{figure}
\begin{tabular}{cc}
\includegraphics[width=.4\columnwidth]{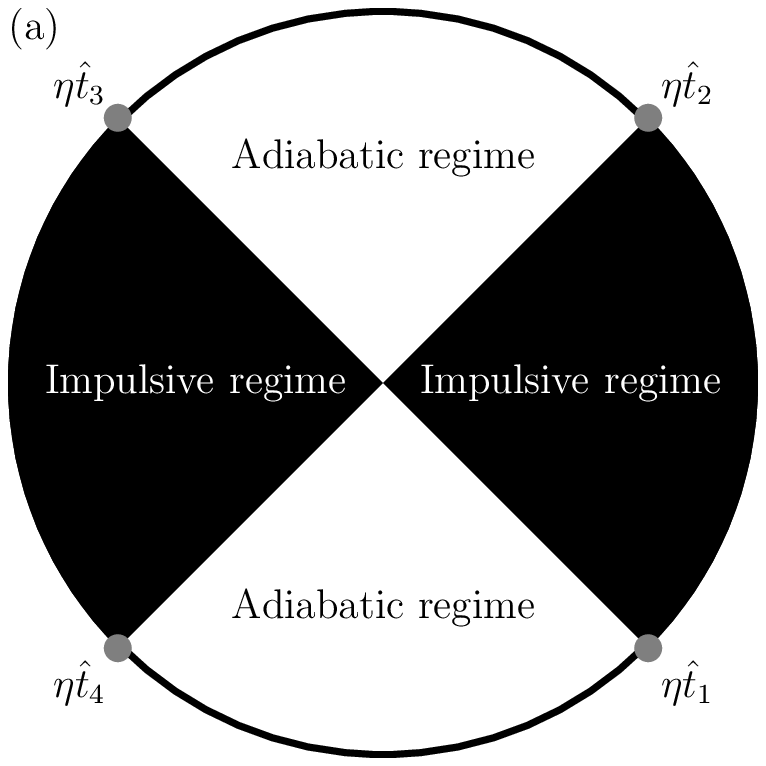}&
\includegraphics[width=.56\columnwidth]{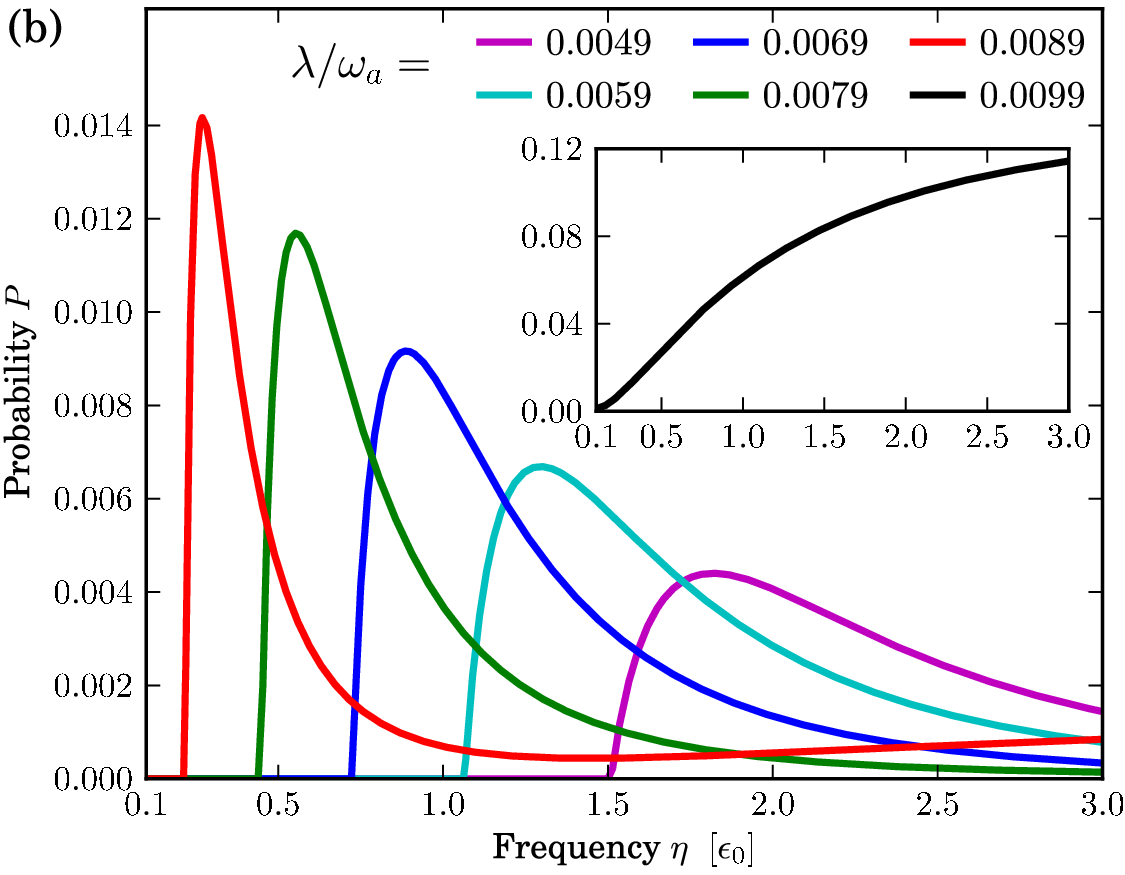}
\end{tabular}
\caption{{\bf } (a) Schematic representation of the four freeze-out points in the trigonometric circle. (b) Probability of leaving the ground state against $\eta/\epsilon_0$ for $g =0.49/\omega_a$ and various values of $\lambda.$  }
\label{KZfig}
\end{figure}

Finally, we address the crucial connection between our DCE-like mechanism and the Kibble-Zurek one~\cite{KZ1,KZ2}.  On approaching the critical 
point of the model in Eq.~(\ref{H0j}), regardless of the value of $\eta$, there will always be a regime where the perturbation is non-adiabatic and photons are produced.   A first estimate of the unavoidable departure from adiabaticity, with a consequent photon-flux, is obtained by calculating the probability of the system to go into an excited state. For simplicity, we consider one period in the absence of damping. The probability of leaving the ground state at the final time $t_f$ ($t_i$ being the initial time) is $P{=}1{-}|\langle \Psi(t_f)|\varphi_0(t_f)\rangle|^2$ with $|\varphi_n(t)\rangle $ the instantaneous eigenstates of the harmonic oscillator and $|\Psi(t_f)\rangle$ the final  state of the system. The KZM relies on the assumption that the state of a  system brought close enough to the critical point {\it freezes} when the system is not able to adiabatically follow the changes in the control parameter. For the driving here at hand, the freeze-out times is found by solving the equation $\mathcal{T}(t)/\dot{\mathcal{T}}(t) = \tau(t),$ where $\mathcal{T}(t) = g_{c}(t)/g -1$ plays the role of the relative temperature of the system and $\tau = \tau_0/\epsilon(t)$  is its relaxation time ($\tau_0 = 1/\omega$)~\cite{Comment}.  For a sinusoidal modulation of  $\mathcal{T}(t)$ and if the oscillating terms brings the system sufficiently close to the critical point, one finds four solutions, each embodying a freeze-out time. Fig.~\ref{KZfig} (a) shows their representations in the unit circle. As the system is initialized in its ground state, i.e. $|\Psi(t_i)\rangle = |\varphi_0(t_i)\rangle,$ the adiabatic condition $\mathcal{T}(t)/\dot{\mathcal{T}}(t) > \tau$ is satisfied until $t=\hat{t}_1,$ where $\hat{t}_1$ is the freeze-out time at which the system enters the so-called impulsive regime. During this period, the state of the system is frozen until $t=\hat{t}_2,$ when the adiabatic condition is fulfilled again and the state of the system becomes $|\Psi(\hat{t}_2)\rangle = |\varphi_0 (\hat{t}_1)\rangle = \sum_n c_{n,0}(\hat{t}_2,\hat{t}_1) |\varphi_n(\hat{t}_2)\rangle$ with $c_{n,m}(t,t^\prime) = \langle \varphi_n(t) | \varphi_m (t^\prime)\rangle.$ The same argument applies to the second part of the cycle, where 
the system  evolves adiabatically  for $t \in [\hat{t}_2,\hat{t}_3]$ and is frozen   for  $t \in [\hat{t}_3,\hat{t}_4].$  The state at $\hat{t}_4$ is then $|\Psi(\hat{t}_4)\rangle = \sum_{k,n} c_{k,n}(\hat{t}_4,\hat{t}_3) c_{n,0}(\hat{t}_2,\hat{t}_1) e^{-i \theta_n}  |\varphi_k(\hat{t}_4)\rangle$ with $\theta_n = \int_{\hat{t}_2}^{\hat{t}_3} dt E_n(t) .$ Finally, the last part of the evolution ($t \in [\hat{t}_4,t_f]$) will not affect the probability $P,$ which is thus $P=1- |\langle \Psi(\hat{t}_4)|\varphi_0(\hat{t}_4)\rangle|^2$ and whose behavior against $\eta$ is shown in Fig.~\ref{KZfig} (b) for different values of $\lambda$. Clearly,  the closer the system to the quantum phase transition, the more it is susceptible to a low-frequency driving. A more detailed analysis requires the study of the transient dynamics. The  scheme of Fig.~\ref{KZfig} (a) is still valid, the probablity of excitations being calculated by composing four different dissipative maps in the same spirit of Ref.~\cite{Shytov}. We only expect quantitative changes.

\begin{figure}[t]
\includegraphics[width=0.6\columnwidth]{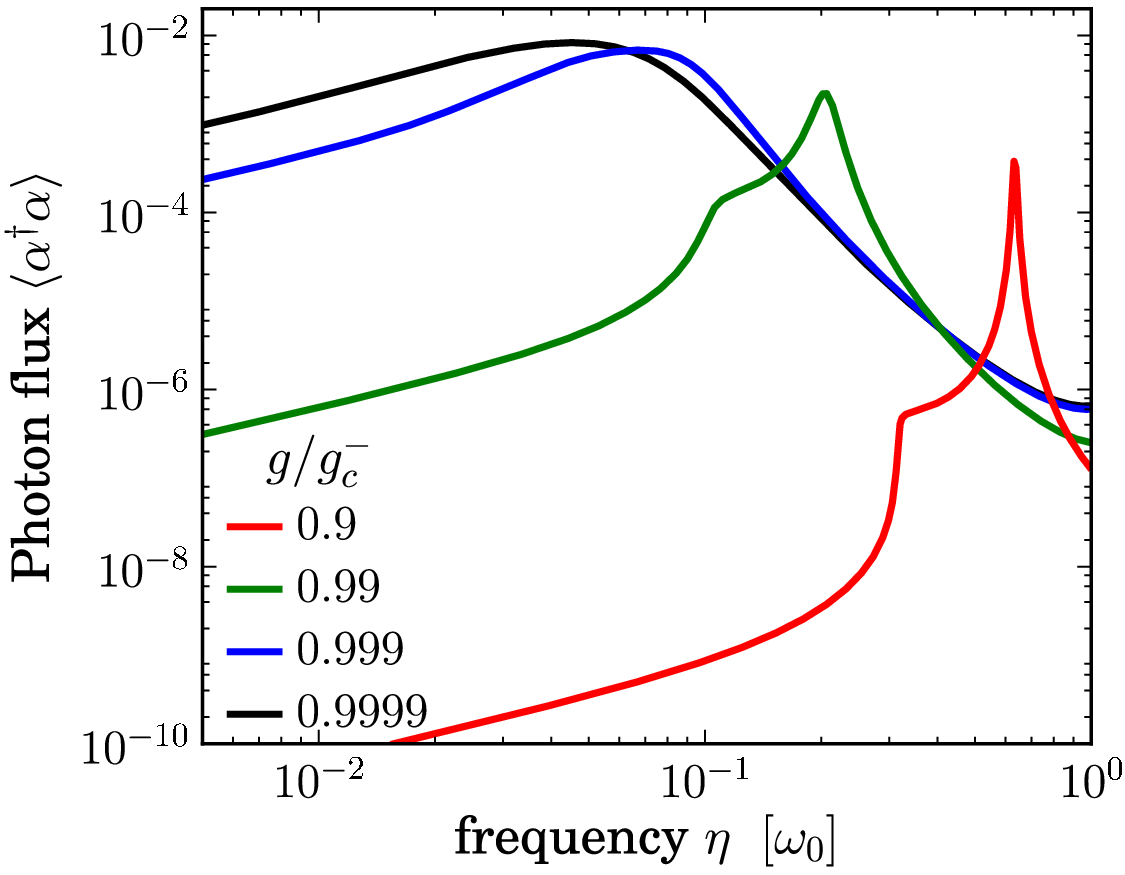}
\caption{Output photon-flux as a function of $\eta$ for different values of $g$. The transition between adiabatic and non-adiabatic regime (sharp step) is located at the minimum of the gap and is shifted to lower frequency when the coupling gets closer to the critical coupling. At the critical point the dynamics is purely non-adiabatic.}
\label{scaling}
\end{figure}

To corroborate the connection between DCE and KZM, we have further analyzed the photon production in the adiabatic and non-adiabatic regimes [cf. Fig.~\ref{scaling}]. For $\eta{>}\epsilon_\mathrm{min}$ (being $\epsilon_\mathrm{min}$ the minimum value of $\epsilon(t)$ over a cycle), the dynamics is non-adiabatic and photons can be created. Close to criticality, the minimum of the gap vanishes, the system is always in the non-adiabatic regime and the photon-flux increases linearly with $\eta$ until the maximum value at resonance is reached. 
Far from the transition, the photon production decreases from the resonance with a Lorentzian behavior: when $\eta{<}\epsilon_\mathrm{min},$ the photon-flux is sharply reduced and a linear behavior is recovered but with a much smaller value. This abrupt transition between the adiabatic and non-adiabatic regimes demonstrates that the breakdown of adiabaticity due to critical slowing-down is at the origin of photon creation in the DCE, a situation totally analogous to what is described by the KZM.

We have proposed a scheme to achieve DCE-like effects by dragging a driven quantum Dicke model across its critical point. By linearizing the model, we have related the mechanism of photon generation from the vacuum to the properties of the eigen-modes of the system, thus providing a clear picture of the DCE-like effect arising from a Dicke quantum phase transition guided by a time-modulated driving. As the frequency of the driving is quenched at the Dicke critical point, the observation of a sizable flux of generated photons becomes less demanding for an optical-domain verification. We have also connected the photon-generation process to the KZ predictions for defect nucleation at a quantum critical point, thus pointing out the intimate connection among three fundamental mechanisms in quantum many-body physics.

 We acknowledge financial support by  the National Research Foundation and Ministry of Education in Singapore, the EU FP7 Programme under Grant Agreement No. 234970-NANOCTM and No.~248629-SOLID, UK EPSRC (EP/G004759/1), and EUROTECH.

\section{Appendix A: Frequencies of the normal modes}

In this Section we provide additional details on the achievement of the two-oscillator model used in the main Letter and give the explicit form of the normal-mode frequencies. We start referring again to the Hamiltonian of the system in the dipole approximation reads (we set $\hbar=1$)
   \begin{equation}
 \hat  H_0= \omega_a \hat{a}^\dag \hat{a} + \omega_b(t) \hat{J}_z + \frac{g}{\sqrt{2j}} (\hat{a}^\dag + \hat{a}) (\hat{J}_+ + \hat{J}_-),
   \label{H0j}
   \end{equation}
where, as discussed in the main Letter, $\omega_a$ is the frequency of the cavity and $g$ is the atom-field coupling constant. For $N\gg{1}$, {the use of the Holstein-Primakoff picture} is well motivated. The angular momentum describing the atomic cloud is thus reduced to a non-linear harmonic oscillator as 
\begin{equation}
\begin{aligned}
&\hat{J}_+{=}\hat{b}^\dag \sqrt{2j ( 1 -\hat{b}^\dag \hat{b}/2j)},\\ 
&\hat{J}_z{=}(\hat{b}^\dag \hat{b}-j).
\end{aligned}
\end{equation}
Here, $\hat b$ and $\hat b^\dag$ are the operators of the effective non-linear boson.
As the number of atoms is large, $j \gg 1$ and a good approximation is given by taking $\hat{J}_+=J^\dag_- \approx \sqrt{2 j} \hat{b}^\dag$. The Hamiltonian thus reduces to Eq. (2)  of the main Letter, which is straightforwardly diagonalized to give the eigen-frequencies
   \begin{equation}
   \epsilon_{\pm}^2(t){=}\frac{1}{2}\left[\omega_a^2 {+} \omega_b^2(t) {\pm} \sqrt{(\omega_b^2(t){-}\omega_a^2)^2{+}16 g^2 \omega_a \omega_b(t)}\right].
   \end{equation}
These are associated with the eigen-modes $\hat q_\pm(t)$ whose explicit form is not relevant for our analysis.

\section{Appendix B: Loss-affected dynamics, out-coming photon flux and spectrum}

In the losses-affected scenario, the dynamics of the cavity mode is conveniently tackled by means of the input-output formalism for optical cavities~\cite{Walls} and by  
modelling the bath as a continuous distribution of harmonic oscillators with frequencies $\nu$, formally described by the bosonic operators $\hat \alpha_{\nu}$ and $\hat \alpha_{\nu}^\dag$ and each coupled to the cavity field with a strength $k_\nu$. As in Eq. (3) of the main Letter, the cavity field-bath coupling Hamiltonian is $V {=} \int_{0}^{\infty} \mathrm{d} \nu k_{\nu} (\hat \alpha_{\nu} \hat a^\dag - \hat a \hat\alpha_{\nu}^\dag )$. The dynamics of the system is analysed by studying the Langevin equations~\cite{Walls}, which can be cast in a compact matrix form. Taking into account that the effective atomic mode $b$ does not experience dissipation for a fully condensed gas without atom losses, we define the bosonic operators vector $\hat u(t)=(\hat a(t),\hat b(t),\hat a^\dag(t),\hat b^\dag(t))^T$ and the Langevin-force vector $\hat F(t)=(\hat f(t),0,\hat f^\dag(t),0)^T$, such that the equations of motion for the atomic and cavity-field modes read 
   \begin{equation}
   \dot{\hat u}(t) = -i M(t) \cdot \hat u(t) - \int \mathrm{d}t^\prime \Gamma(t-t^\prime) \hat u(t^\prime) + \hat F(t).
   \label{langevin2}
   \end{equation}
Here $M(t)$ is a time-dependent $4\times4$ matrix that takes into account the unitary evolution and $\Gamma(t{-}t^\prime){=} \text{diag}[\gamma(t{-}t^\prime),0,\gamma(t{-}t^\prime),0]$ is the dissipation kernel whose elements $\gamma(t-t')$ are decay rates. The explicit form of such rates will be discussed later on. 
In light of our choice for the modulation of the atomic energy splitting, the elements of $M(t)$ are oscillating functions with period $T=2 \pi/\eta$, so that we can take $M(t) = M_0 + M_1 (e^{i\eta t} {-} e^{-i \eta t})$ with $M_0$ the time-independent term 
   \begin{equation}
   M_0 =\begin{pmatrix}
   \omega_a & g &0&g\\
   g & \omega_0 & g & 0\\
   0 & -g & - \omega_a & - g \\
   -g & 0 & -g & -\omega_0 
   \end{pmatrix},
   \label{matrixM0} 
   \end{equation} 
and $M_1 = \text{diag}[0,\lambda,0,-\lambda]$ that takes into account the modulation. 
Notice that, due to the inclusion of the counter-rotating terms, the decay rates in $\Gamma(t-t^\prime)$ in Eq.~(\ref{langevin2}) are time-dependent. Solving Eq.~(\ref{langevin2}) is made difficult, in general, by the convolution integral describing damping. The approach is significantly simplified by moving to the frequency domain. The Langevin-force operators $\hat {f}(\omega)$ in the frequency domain are linked to the input noise operators as ${\hat f}(\omega)= 2 \pi k_{\omega} \rho(\omega) \hat \alpha^\mathrm{in}_{\omega}$, where $\rho(\omega)$ is the density of states of the bath. The decay rates $\gamma(\omega)\in\mathbb{C}$ are thus found to have ${\Re}[\tilde{\gamma}(\omega)] = \pi |k_{\omega}|^2 \rho(\omega)$ and ${\Im}[\tilde{\gamma}(\omega)]= -\frac{1}{\pi} {\cal P}\int_{-\infty}^{\infty} \mathrm{d}\omega^\prime {\Re}[\tilde{\gamma}(\omega^\prime)]/(\omega^\prime - \omega)$, where ${\cal P}$ denotes the principal value of the integral~\cite{CiutiCaru}. 
While the imaginary part of $\tilde{\gamma}(\omega)$ is just a frequency shift, ${\Re}[\tilde{\gamma}(\omega)]$ is responsible for the frequency-dependent damping of the cavity mode. 
When the counter-rotating terms are taken into account, it becomes crucial to consider that the density of photonic state in the bath $\rho(\omega)$ is zero for negative frequencies. 
Indeed, by modeling the bath as a collection of harmonic oscillators, only positive frequencies have a physical meaning. 
Within this assumption, it follows immediately from the definitions given above that $\tilde{\gamma}(\omega) = 0$ and $\tilde{f}(\omega) = 0$ for $\omega < 0$. 
We will also suppose that the damping rate is constant for positive frequencies. 
This is equivalent to assuming that $k_{\omega}= k$ and, for $\omega>0$, $\rho(\omega) = 1$ Within this assumption, we define $\gamma_0 \equiv \pi |k|^2$ and we can write $\text{Re}[\tilde{\gamma}(\omega>0)] = \gamma_0$ and $\text{Re}[\tilde{\gamma}(\omega<0)] = 0$.

In the following the index $m=0,\pm1$ indicates the number of sidebands in the drive and $\mathcal{G}_{i,j}(\omega)$ are the matrix elements of $\mathcal{G}(\omega)= [i\mathcal{M}(\omega)]^{-1}$. 
We first consider the mean number of photons inside the cavity at the stationary state $\langle \hat a^\dag \hat a \rangle= \lim_{t\rightarrow \infty} \langle \hat a^\dag(t) \hat a(t) \rangle$.
After Fourier transforming $\hat a(t)$ and in the limit $t{\rightarrow}\infty$ we have $\langle \hat a^\dag \hat a \rangle{=}(2\pi)^{-2} \int_{-\infty}^{\infty} \mathrm{d}\omega \langle \tilde{a}^\dag(\omega) \tilde{a}(\omega) \rangle$. Using the steady state solution of the cavity operator and the expression for $\tilde{f}(\omega)$ given in the last paragraph,  $\tilde{f}(\omega)= 2 \pi k_{\omega} \rho(\omega) \alpha^\mathrm{in}_{\omega}$, the mean number of photons inside the cavity at the steady state is given by 
   \begin{equation}
   \langle \hat a^\dag \hat a \rangle_{(m)} = \frac{\gamma_0}{\pi} \sum_{j=-m}^{m} \int_{j\eta}^{\infty} \mathrm{d}\omega |\mathcal{G}_{4m+1,4(m+j)+3}(-\omega)|^2. 
   \label{photonsinside}
   \end{equation}
Here we have assumed that the input field is in the vacuum state, so the operators $\hat\alpha^\mathrm{in}_{\omega}$ fulfill the condition $\langle\hat \alpha^\mathrm{in}_{\omega} \hat\alpha^{\mathrm{in}^\dag}_{\omega^\prime} \rangle = \delta(\omega - \omega^\prime)$. A factor of $\rho(-\omega - j \eta)$ is responsible for the shift in the integration limits. 

In order to obtain the output operator of the cavity $\hat\alpha^\mathrm{out}_{\omega}$, we substitute the solution for the cavity operator $\tilde{a}(\omega)$ into the input-output relation~\cite{CiutiCaru}. 
Doing so, the output operator reads as
   \begin{multline}
   \hat\alpha_{\omega}^\mathrm{out} = \hat\alpha_{\omega}^\mathrm{in} - 2\gamma_0 \sum_{j=-m}^{m} \mathcal{G}_{4m+1,4(m+j)+1}(\omega) \rho(\omega +j\eta) \hat\alpha_{\omega + j\eta}^\mathrm{in} \\ 
   + 2\gamma_0 \sum_{j=-m}^{m} \mathcal{G}_{4m+1,4(m+j)+3}(\omega) \rho(-\omega - j\eta) \hat\alpha_{-\omega - j\eta}^{\mathrm{in}^\dag}.
   \label{alphaout}
   \end{multline}
The expression for the photonic flux outside the cavity is $\langle\hat \alpha^\dag \hat\alpha \rangle = \int_{0}^{\infty} \mathrm{d}\omega \langle \hat\alpha^{\mathrm{out}^\dag}_{\omega} \hat\alpha^\mathrm{out}_{\omega} \rangle$. {Since $\rho(\omega) = 0$ for $\omega < 0$   the only non-vanishing terms in the summation are the ones with $j<0$. }
Notice that the negative frequencies have been taken out from the integration since only positive frequencies are physically allowed in the output field. 
Having the matrix elements of $\mathcal{G}$ the dimension of time, the quantity $\langle \hat\alpha^\dag \hat\alpha\rangle$ has the correct dimension of $1/t$ for a flux of photons. 

Finally, in order to calculate the spectral density $S(\omega)$ of out-coming photons, we first evaluate the analogous quantity $\mathcal{P}(\omega)=\langle a^\dag(\omega) a(\omega)\rangle$ inside the cavity. From this we find $\mathcal{S}(\omega)$ using the input-output relations, with the result $\mathcal{S}(\omega) =( \gamma_0/\pi ) \mathcal{P}(\omega)$.


\begin{thebibliography}{99}
\bibitem{Comment} 
	Notice that $\mathcal{T}(t)$ is defined as $\mathcal{T}(t) = 1-g(t)/g_c$ if the time-dependent parameter is $g$ instead of $\omega.$
\end{thebibliography}
\end{document}